\documentclass[aps,prl,showpacs,twocolumn,superscriptaddress]{revtex4-1}

\usepackage{amsmath}
\usepackage{amsfonts}
\usepackage{amssymb}
\usepackage{epsfig}

\usepackage{braket}

\usepackage{graphicx}

\usepackage{bbold}

\usepackage{hyphenat}
\usepackage{hyperref}
\usepackage{placeins}	
\usepackage{color}

\usepackage{array}
\newcolumntype{C}[1]{>{\centering\arraybackslash}m{#1}}
\usepackage{overpic}

\usepackage[normalem]{ulem}

\usepackage{booktabs}
\usepackage{diagbox}
\usepackage{siunitx}
\AtBeginDocument{
	\heavyrulewidth=.08em
	\lightrulewidth=.05em
	\cmidrulewidth=.03em
	\belowrulesep=.65ex
	\belowbottomsep=0pt
	\aboverulesep=.4ex
	\abovetopsep=0pt
	\cmidrulesep=\doublerulesep
	\cmidrulekern=.5em
	\defaultaddspace=.5em
}
\begin{document}

\title{Fracton and topological order in the XY checkerboard toric code}
\author{Maximilian Vieweg}
\affiliation{Department Physik, Staudtstra{\ss}e 7, Universit\"at Erlangen-N\"urnberg, D-91058 Erlangen, Germany}

\author{Kai Phillip Schmidt}
\affiliation{Department Physik, Staudtstra{\ss}e 7, Universit\"at Erlangen-N\"urnberg, D-91058 Erlangen, Germany}

\begin{abstract}
We introduce the XY checkerboard toric code. It represents a generalization of the $\mathbb{Z}_2$ toric code with two types of star operators with $x$ and $y$ flavor and two anisotropic star sublattices forming a checkerboard lattice. The quantum phase diagram is deduced exactly by a duality transformation to two copies of self-dual Xu-Moore models, which builds on the existence of an sub-extensive number of $\mathbb{Z}_2$ conserved parities. 
For any spatial anisotropy of the star sublattices, the XY checkerboard toric code realizes two quantum phases with $\mathbb{Z}_2$ topological order and an intermediate phase with type-I fracton order. The properties of the fracton phase like sub-extensive ground-state degeneracy can be analytically deduced from the degenerate limit of isolated stars. For the spatially isotropic case the extension of the fracton phase vanishes. The topological phase displays anyonic excitations with restricted mobility and dimensional reduction. All phase transitions are first order. 
\end{abstract}

\maketitle

Two-dimensional quantum matter with topological order \cite{Wen_1989,Wen_1990,Wen_2004} exhibits highly entangled ground states and fractional excitations with unconventional anyonic statistics \cite{Leinaas_1977,Wilczek_1982}. These fascinating quantum phases are proposed for potential applications in quantum technologies, particularly in topological quantum computation or quantum memories \cite{Kitaev_2003,Nayak_2008}. On the experimental side, topological phases are investigated in fractional quantum Hall systems \cite{Laughlin_1983,Tsui_1982} and frustrated quantum magnetism \cite{Balents_2010,Jackeli_2010,Singh_2012,Plumb_2014,Banerjee_2017,Savary_2017,Banerjee_2018} in condensed matter physics as well as with quantum simulators utilizing trapped ions \cite{Han_2007}, photons \cite{Lu_2009,Pachos_2009} or NMR \cite{Du_2007,Feng_2012,Peng_2014} and are proposed for several other quantum-optical platforms \cite{Micheli_2006,Paredes_2008,Sameti_2017}.

The concept of topological order can be extended to three dimensions \cite{Chamon_2005,Bravyi_2011,Haah_2011,Yoshida_2013,Vijay_2015,Vijay_2016}.  
While some properties like the ground-state degeneracy depending on topology are similar, 
point-like anyons are forbidden so that
non-trivial statistics can only be found for extended excitations. 
Further, one must distinguish between these long-range entangled ground states 
and fracton topological order \cite{Chamon_2005,Bravyi_2011,Haah_2011,Yoshida_2013,Vijay_2015,Vijay_2016} with sub-extensive ground-state degeneracy.
One essential property of fracton phases is that their elementary excitations have a
restricted mobility under the action of local operators.
As a consequence, fracton phases of matter are connected to glassy dynamics and spin liquids \cite{Nandkishore2019}. 
Key aspects like sub-extensive ground-state degeneracy and kinetic constraints can naturally be also
present in two-dimensional fracton phases resulting from subsystem symmetries \cite{Xu_2004,Xu_2005,Newman1999,Vasiloiu2020,Wiedmann2024}.

The paradigmatic model for topological order is Kitaev's exactly solvable two-dimensional $\mathbb{Z}_2$ toric code \cite{Kitaev_2003},  which has been proposed as quantum memory and is relevant for quantum error correction. The toric code has a $\mathbb{Z}_2$ topologically ordered ground state and a genus-dependent ground-state degeneracy. Its elementary excitations are mutual Abelian anyons. The analytic solution triggered many investigations about the physical properties of topological order, e.g., its robustness on the quantum \cite{Trebst_2007,Hamma_2008_b,Yu_2008,Vidal_2009,Vidal_2011,Dusuel_2009,Tupitsyn_2010,Wu_2012,Dusuel_2011,Schmidt_2013,Morampudi_2014,Zhang_2017,Vanderstraeten_2017} and thermal \cite{Alicki_2009,Castelnovo_2007,Nussinov_2009_b} level. 
Apart from changing the underlying symmetry group \cite{Kitaev_2003,Bais_2009,WOOTTON20112307,Bullock2007,Schulz_2012}, generalizations of the toric code are invented by applying combinatorial gauge symmetry \cite{Chamon_2020}, which can be implemented in superconducting quantum circuits \cite{Chamon_2021}. 
This has further led to the analytically non-solvable $U(1)$-symmetry enriched toric code \cite{Wu_2023}. 
For the latter, numerical evidence was found for a topologically ordered gapped quantum spin liquid with various peculiar properties like UV/IR mixing and Hilbert space fragmentation \cite{Wu_2023}.

\begin{figure}[t]
       \centering
        \includegraphics[width=0.9\columnwidth]{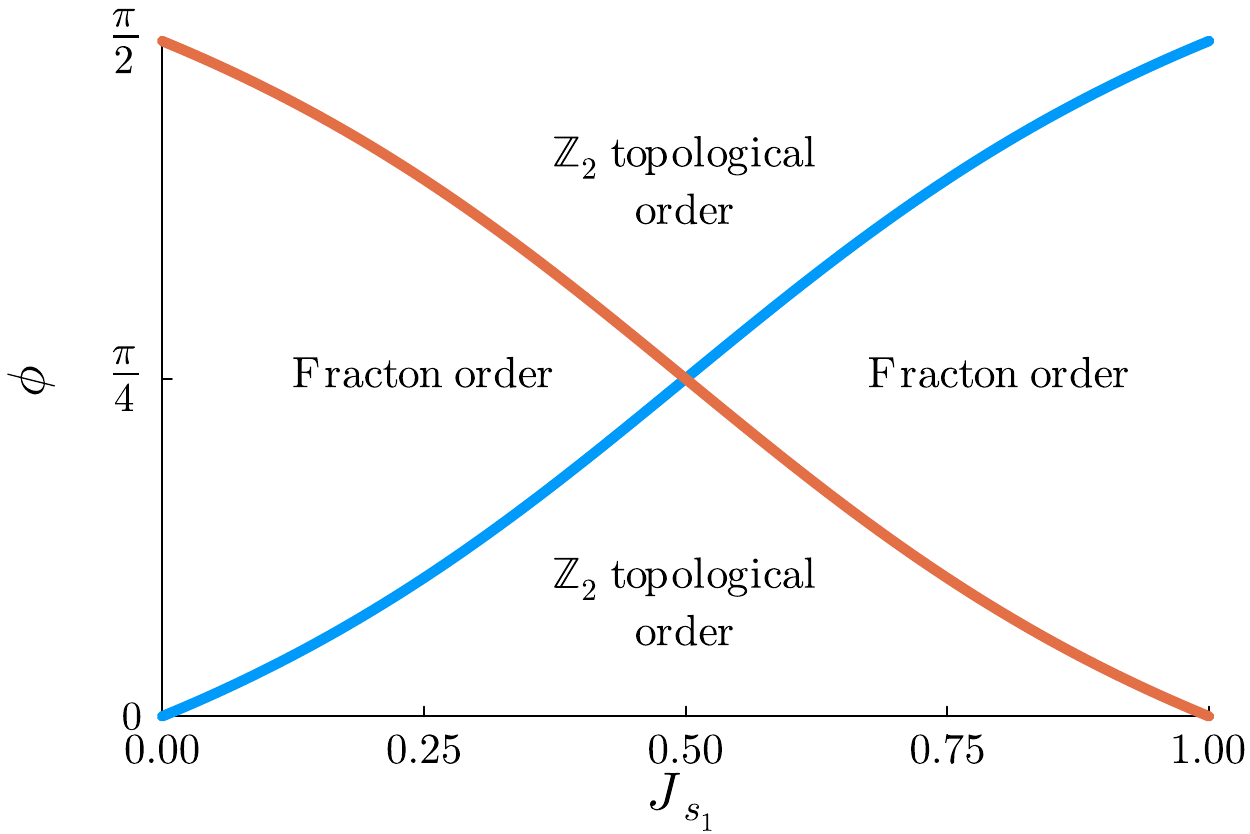}
        \caption{Quantum phase diagram of the XYTC as a function of $J_{s_1}$ with $J_{s_2}=1-J_{s_1}$ (spatial anisotropy between the two star sublattices) and the XY anisotropy $\phi$ of star operators. The red and blue solid lines represent the first-order phase transition lines $J_{s_1}\cos(\phi)=J_{s_2}\sin(\phi)$ and $J_{s_2}\cos(\phi)=J_{s_1}\sin(\phi)$ separating topological and fracton order. The diagram is symmetric under reflections about $\phi=\pi/4$ and $J_{s_1}=1/2$. 
        }
       \label{Fig::XYTC_pd}
\end{figure}

In this letter we 
introduce the XY checkerboard toric code (XYTC) connecting topological and fracton order in two dimensions. 
Related to both, the conventional $\mathbb{Z}_2$ toric code and its $U(1)$ enriched version, the XYTC generalizes the toric code to two types of star operators with $x$ and $y$ flavor and two anisotropic star sublattices resulting in a checkerboard lattice. 
In contrast to the $U(1)$ toric code \cite{Wu_2023}, an exact mapping and analytic considerations allow us to quantitatively determine the quantum phase diagram shown in Fig.~\ref{Fig::XYTC_pd} establishing the presence of topological and fracton order.  

\begin{figure}[t]
        \centering
        \includegraphics[width=0.8\columnwidth]{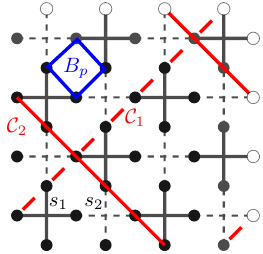}
        \caption{Sketch of the XYTC with linear extension $L=4$ and periodic boundary conditions. Circles represent spins-1/2 located on the links of the checkerboard lattice. Star sublattices $s_1$ and $s_2$ are illustrated as thick and thin gray crosses. An exemplary plaquette $p$ is shown in blue. Solid (dashed) red line represents an exemplary diagonal (anti-diagonal) contour $\mathcal{C}_1$ ($\mathcal{C}_2$) on which conserved subsystem parity operators $\prod_{i\in\mathcal{C}}\sigma_i^x$ are defined.}
        \label{Fig::XYTC}
\end{figure}

{\it{Model:}}
The XYTC on a checkerboard lattice is illustrated in Fig.~\ref{Fig::XYTC}. 
Its Hamiltonian is given by
\begin{eqnarray}
 \mathcal{H}_{\rm XYTC}&=&-\sum_{p}\hat{B}_{p}\nonumber\\
                       && -J_{s_1}\sum_{s_1}\left( \cos(\phi)\hat{A}_{s_1}^{(x)}+\sin(\phi)\hat{A}_{s_1}^{(y)}\right)\nonumber\\
                       && -J_{s_2}\sum_{s_2}\left( \cos(\phi)\hat{A}_{s_2}^{(x)}+\sin(\phi)\hat{A}_{s_2}^{(y)}\right)\,,\label{eq:xytc}
\end{eqnarray}
with plaquette and star operators $\hat{B}_{p}=\prod_{i\in p}\sigma_{i}^z$, \mbox{$\hat{A}_{s}^{(x)}=\prod_{i\in s}\sigma_{i}^x$}, and $\hat{A}_{s}^{(y)}=\prod_{i\in s}\sigma_{i}^y$ having eigenvalues $b_{p},a_{s}^{(x)},a_{s}^{(y)}\in\{\pm 1\}$, respectively. 
The unit cell of the XYTC is taken to be the four sites of stars $s_1$ which build an effective square lattice rotated by $45^\circ$. 
The linear extension of the lattice is called $L$.
In the following we consider $J_{s_{2}}=1-J_{s_{1}}$ with $J_{s_{1}}\in [0,1]$ due to the symmetry between the two star sublattices and $\phi\in [0,\pi/2]$ due to the symmetry between $x$ and $y$ star operators. The XYTC reduces to the $\mathbb{Z}_2$ toric code 
with topological order for the cases $\phi =0$ and $\phi=\pi/2$.
Let us remark that the isotropic case $J_{s_{1}}=J_{s_{2}}$ and $\phi=\pi/4$ of the XYTC is the same as the $U(1)$ toric code \cite{Wu_2023} up to a term proportional to $(\prod_{i\in s}\sigma_{i}^+ + \prod_{i\in s}\sigma_{i}^- )$ on each star.  

{\it{Conserved quantities:}} 
The individual plaquette operators $\hat{B}_{p}$ commute with $\mathcal{H}_{\rm XYTC}$ so that the $b_{p}=\pm 1$ eigenvalues are conserved quantities. 
This is not the case for $a_{s}^{(x)}$ and $a_{s}^{(y)}$ because \mbox{$[\hat{A}_{s}^{(x)},\hat{A}_{s}^{(y)}]\neq 0$}.  
Most importantly, the subsystem parity operator $\prod_{i\in\mathcal{C}}\sigma_i^x$ commutes with $\mathcal{H}_{\rm XYTC}$, provided $\mathcal{C}$ is a diagonal or anti-diagonal contour such as the ones depicted in Fig.~\ref{Fig::XYTC}. 
In addition, the string operators $\prod_{i\in\mathcal{C}_{\rm nc}}\sigma_i^z$ with $\mathcal{C}_{\rm nc}$ non-contractible loops of the underlying topology with genus $g$ commute with $\mathcal{H}_{\rm XYTC}$ as for the conventional toric code. However, we stress that string and subsystem parity operators are not independent.

As for the toric code in a magnetic field in $y$-direction \cite{Vidal_2011}, the conserved subsystem parities have tremendous consequences on the physical properties of the XYTC, in particular the dynamics of excitations is strongly constraint. 
Physically, in the $x$-basis, the subsystem parities of the number of spin flips along any diagonal and anti-diagonal is thus conserved.
Let us note that also the generalized subsystem parity operators \mbox{$\alpha\prod_{i\in\mathcal{C}}\sigma_i^x + \beta\prod_{i\in\mathcal{C}}\sigma_i^y$}
 with $\alpha,\beta$ constant commute with $\mathcal{H}_{\rm XYTC}$, but do not result in independent conserved quantities so that we have chosen $\alpha=1$ and $\beta=0$. 
 
{\it{Duality mapping:}} The XYTC is isospectral to two copies of Xu-Moore (XM) models \cite{Xu_2004,Xu_2005} in the thermodynamic limit with open boundary conditions. 
The mapping does not keep track of degeneracies.
As in Ref.~\cite{Vidal_2011}, we introduce pseudo-spin 1/2 variables living on the dual lattice so that 
\begin{align}\label{eq:duality}
    \tau^z_{j_s} = \hat{A}_s^{x}\,,\quad \tau^z_{j_p} = \hat{B}_p\,,\quad {\text{and}}\quad \tau^x_{j} = \prod_{j>i}\sigma^y_i\,,
\end{align}
where $j_{s(p)}$ denotes the center of an $x$-star (plaquette).
The notation $j>i$ defines the set of all spin sites $i$ whose two coordinates are smaller than those of $j$ where $j$ denotes the center of stars and plaquettes.
The XYTC can then be rewritten as 
\begin{align}\label{eq:xytcdual}
 \mathcal{H}_{\rm dual}=-\sum_{j_p}\tau^z_{j_p}+\mathcal{H}_{\rm XM}^{(s_1)}+\mathcal{H}_{\rm XM}^{(s_2)}
\end{align}
with two kind of XM models 
\begin{eqnarray}
      \mathcal{H}_{\rm XM}^{(s_1)} &=& -J_{s_1}\cos(\phi)\sum_{j_{s_1}}\tau^z_{j_{s_1}}-J_{s_2}\sin(\phi)\sum_{\tilde{p}_1}\prod_{j_{s_1}\in\tilde{p}_1}\tau^x_{j_{s_1}}\nonumber\\
       \mathcal{H}_{\rm XM}^{(s_2)} &=&  -J_{s_2}\cos(\phi)\sum_{j_{s_2}}\tau^z_{j_{s_2}} -J_{s_1}\sin(\phi)\sum_{\tilde{p}_2}\prod_{j_{s_2}\in\tilde{p}_2}\tau^x_{j_{s_2}}\,,\nonumber\label{eq:XMl}
\end{eqnarray}
where the first sum runs over all sites $j_{s_1}$ ($j_{s_2}$) of star sublattice $\tilde{\Lambda}_{s_1}$ ($\tilde{\Lambda}_{s_2}$). The second sum is carried over plaquettes $\tilde{p}_{s_1}$ ($\tilde{p}_{s_2}$) which consist of the four center of stars $j_{s_1}$ ($j_{s_2}$) forming a minimal square on star sublattice $\tilde{\Lambda}_{s_1}$ ($\tilde{\Lambda}_{s_2}$).

Obviously, $[\mathcal{H}_{\rm XM}^{(s_1)},\mathcal{H}_{\rm XM}^{(s_2)}]=0$ holds. Each self-dual XM model is known to host two quantum phases with a first-order phase transition at the self-dual point, i.e., $J_{s_1}\cos(\phi)=J_{s_2}\sin(\phi)$ for $\mathcal{H}_{\rm XM}^{(s_1)}$ and \mbox{$J_{s_2}\cos(\phi)=J_{s_1}\sin(\phi)$} for $\mathcal{H}_{\rm XM}^{(s_2)}$. The corresponding ground-state phase diagram of the XYTC consists therefore of two first-order phase transition lines except for the isotropic case $J_{s_1}=J_{s_2}$ where both transition lines cross for $\phi=\pi/4$ (see Fig.~\ref{Fig::XYTC_pd}). The quantum phase adiabatically connected to the limiting toric code cases $\phi=0$ and $\phi=\pi/2$ display $\mathbb{Z}_2$ topological order. The intermediate phases present for $J_{s_1}\neq J_{s_2}$ are type-I fracton phases which we show next by investigating the limit of isolated star sublattices.

\begin{figure}[t]
        \centering
    \includegraphics[width=0.9\columnwidth]{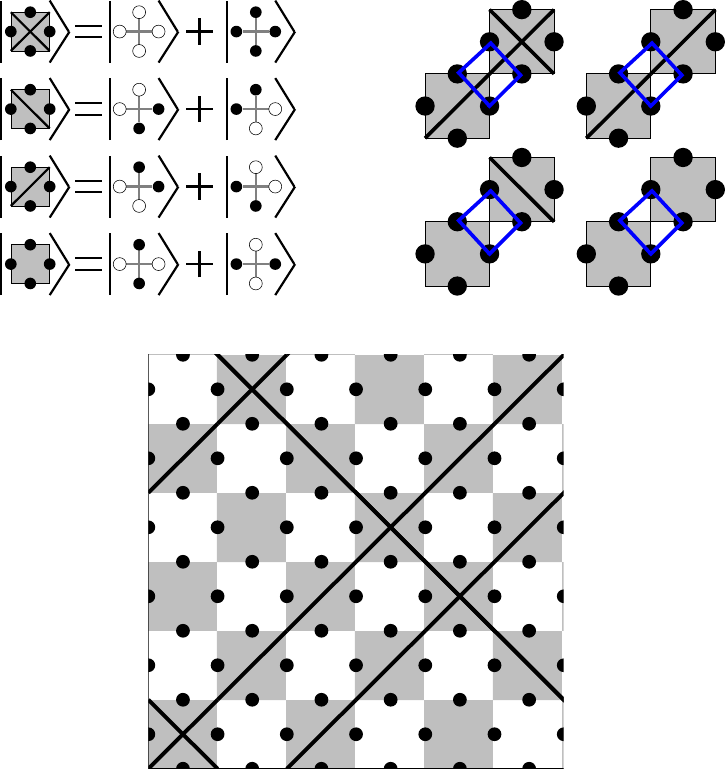}
        \caption{Ground-state manifold for the limit $J_{s_2}=0$ with $\phi\neq\{0,\pi/2\}$ in the subspace with $b_p=+1$ for all $p$. {\it Top left}: Definition and pictorial representation of the four degenerate ground states of a single star $s_1$ (omitting the normalization constant). Empty (filled) circles on the right-hand side refer to spin down (up) in $z$-quantization. {\it Top right}: All possible symmetry-unrelated nearest-neighbor star configurations under the constraint $b_p=1$ on the blue plaquette. {\it Bottom}: Exemplary ground state fulfilling all constraints set by the plaquette operators on a periodic cluster of linear extension $L=6$.}
        \label{Fig::gs}
\end{figure}

{\it{Limit isolated star sublattice:}} We consider the limit $J_{s_2}=0$ of isolated $s_1$ stars. For $\phi\neq\{0,\pi/2\}$, one has four degenerate ground states for each star operator $\cos(\phi)\hat{A}_{s_1}^{(x)}+\sin(\phi)\hat{A}_{s_1}^{(y)}$ with eigenenergy $\cos(\phi) +\sin(\phi)$ resulting in the extensive degeneracy $4^{N_{s_1}}$ with $N_{s_1}$ the total number of $s_1$ stars. In $z$-quantization, these four states correspond to the symmetric combination of zero and four down spins as well as of two down spins oriented  anti-diagonal, diagonal, and vertical/horizontal (see Fig.~\ref{Fig::gs} top left for a pictorial representation).  

This degeneracy is reduced to a sub-extensive level by restricting ourselves to the relevant low-energy subspace with eigenvalues $b_p=+1$ for all $p$. In the following we consider a system of linear extension $L$ and periodic boundary conditions implying $L$ even. Indeed, the condition $b_p=+1$ implies an even number of down spins on every plaquette $p$ so that neighboring star configurations along (anti-) diagonals are not independent anymore. As a result, the configuration of the product state ground states is fully fixed when fixing the local configuration on stars $s_1$ along a single horizontal (or vertical) line. This directly implies a reduction to a sub-extensive ground-state degeneracy scaling only with linear system size.  

Explicitly, we have visualized all possible symmetry-unrelated nearest-neighbor star configurations under the constraint $b_p=1$, where $p$ is the plaquette connecting two $s_1$ stars, in Fig.~\ref{Fig::gs} top right. Enforcing the constraint $b_p=1$ on all plaquettes, one observes that ground states have to be constructed such that diagonals and anti-diagonals of the local configurations have to be fit together, i.e., they are never allowed to possess an open end. This is exemplified in Fig.~\ref{Fig::gs} bottom. As a consequence, the information of a local configuration on a column (or row) of stars $s_1$ is sufficient to determine the associated ground state on the full lattice. The sub-extensive ground-state degeneracy is therefore $2^L$ because one has $4$ local ground-state configurations and $L/2$ number of stars $s_1$ per row (or column). Next we demonstrate that this sub-extensive degeneracy for $J_{s_2}=0$ is stable for small $J_{s_2}$ and that the elementary excitations are fractons and lineons. 
In fact, these properties are robust against any local perturbation as long as the subsystem parity symmetry is respected.
Therefore, the associated phase is a type-I fracton phase. 

{\it{Fracton phase:}} The sub-extensive degeneracy for \mbox{$J_{s_2}=0$} is indeed stable to any order (degenerate) perturbation theory in $J_{s_2}$ in the thermodynamic limit. First, on a finite system with periodic boundary conditions, the minimal perturbative order for which two ground states of the sub-extensive manifold are coupled by $J_{s_2}$ is of order $L$, because one always has to change the local states on stars $s_1$ at least along a whole (anti-) diagonal of the lattice. As a consequence, one only obtains diagonal energy corrections to all ground states in the thermodynamic limit. 

Due to the conserved subsystem parity operators $\prod_{i\in \mathcal{C}}\sigma_{i}^x$, all these energy corrections are identical to any order perturbation theory. This can be seen as follows. Graphically, the action of a single subsystem parity operator $\prod_{i\in \mathcal{C}}\sigma_{i}^x$ introduces (or removes) either a diagonal or anti-diagonal in the pictorial representation (see Fig.~\ref{Fig::gs}) of the local configurations on stars $s_1$ which share spins with $\mathcal{C}$. One therefore obtains a different ground state $\prod_{i\in \mathcal{C}}\sigma_{i}^x\ket{{\rm gs}_i}=\ket{{\rm gs}_j}$ when $\ket{{\rm gs}_i}$ is expressed in the chosen basis. Denoting $\mathcal{H}_{\rm eff}^{(n)}$ the effective Hamiltonian in order $n$ perturbation theory acting in the sub-extensive subspace of ground states, one has
\begin{align}\label{eq:deg}
   \prod_{i\in \mathcal{C}}\sigma_{i}^x \mathcal{H}_{\rm eff}^{(n)} \ket{{\rm gs}_i} =  \mathcal{H}_{\rm eff}^{(n)} \prod_{i\in \mathcal{C}}\sigma_{i}^x  \ket{{\rm gs}_i} = \mathcal{H}_{\rm eff}^{(n)}  \ket{{\rm gs}_j} \,,
\end{align}
and therefore $\ket{{\rm gs}_i}$ and $\ket{{\rm gs}_j}$ remain degenerate.
The same holds for any pair of ground states when acting with several subsystem parity operators.
Consequently, the sub-extensive ground-state degeneracy cannot be broken by small $J_{s_2}$ as well as any local perturbation which respects the subsystem parity symmetry. 
Taking into account the duality mapping shown above, this fracton phase extends up to the phase transition lines towards the two topologically ordered phases (see Fig.~\ref{Fig::XYTC_pd}).

\begin{figure}[t]
    \centering
    \includegraphics[width=0.9\columnwidth]{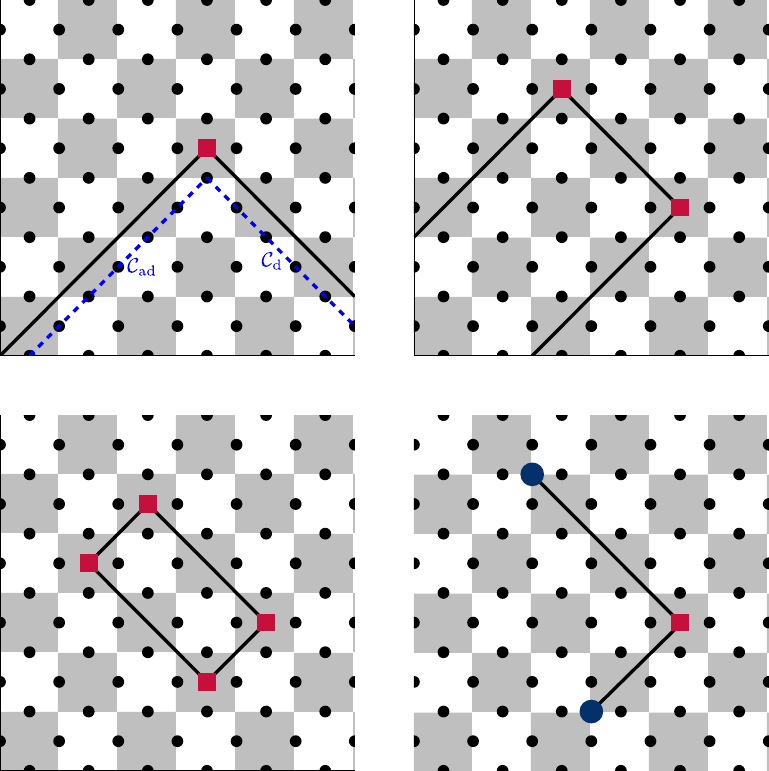}
        \caption{Fracton excitations for $J_{s_2}=0$ with $\phi\neq\{0,\pi/2\}$ in the subspace with $b_p=+1$ for all $p$. {\it Top left}: Single fracton excitation (red square) created on an open plane by $\prod_{i\in\mathcal{C}}\sigma^{x}_i$ or $\prod_{i\in\mathcal{C}}\sigma^{y}_i$ where $\mathcal{C}=\mathcal{C}_{\rm d}\cup\mathcal{C}_{\rm ad}$ and $\mathcal{C}_{\rm d}$ ($\mathcal{C}_{\rm ad}$) semi-infinite strings along the diagonal (anti-diagonal) touching at the fracton star.        
         {\it Top right}: Two-fracton lineon excitation which can only move in one dimension along the two indicated infinite black lines. {\it Bottom left}: Four-fracton excitation on a rectangular with two-dimensional mobility. {\it Bottom right}: Configuration of two plaquette excitations with $b_p=-1$ (blue circles) and a single fracton. 
        }
        \label{Fig::fractons}
\end{figure}

Now we consider the low-energy excitations of the fracton phase which can either be on stars $s_1$ or plaquettes $p$. For general $\phi$ the local spectrum on each star operator is given by eigenvalues $\cos(\phi)+\sin(\phi)$, $\pm \cos(\phi)\mp \sin(\phi)$, and $-\cos(\phi)-\sin(\phi)$.  Elementary excitations correspond up to rotational symmetry to eigenvectors 
$(\ket{\includegraphics[width=.03\linewidth]{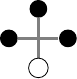}}+\ket{\includegraphics[width=.03\linewidth]{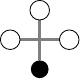}})$ and $(\ket{\includegraphics[width=.03\linewidth]{./config1.pdf}}-\ket{\includegraphics[width=.03\linewidth]{./config2.pdf}})$
with eigenvalues $\pm\cos(\phi)\mp\sin(\phi)$ , respectively.  
Such star excitations can be created on an open plane by $\prod_{i\in\mathcal{C}}\sigma^{x}_i$ or $\prod_{i\in\mathcal{C}}\sigma^{y}_i$ where $\mathcal{C}=\mathcal{C}_{\rm d}\cup\mathcal{C}_{\rm ad}$ and $\mathcal{C}_{\rm d}$ ($\mathcal{C}_{\rm ad}$) semi-infinite strings along the diagonal (anti-diagonal) crossing at a specific site $j$ as illustrated in Fig.~\ref{Fig::fractons} top left. These star excitations are type-1 fractons in the sense, that an isolated fracton cannot be moved by any local perturbation without creating other excitations. 
Indeed, to move a single fracton, one has to change the local ground-state configuration of stars $s_1$ along a whole (anti-) diagonal which is impossible by any operator with a finite extent.  
Along the same lines, pairs of fractons are lineons being only mobile in one dimension along diagonals or anti-diagonals (see Fig.~\ref{Fig::fractons} top right),  
while four-fracton configurations have a two-dimensional dispersion when placed on a rectangular as shown in Fig.~\ref{Fig::fractons} bottom left.
In contrast, the local high-energy star excitation with eigenvalue $-(\cos(\phi)+\sin(\phi))$, which can be created locally by $\sigma^{z}_j$, do not have a restricted mobility.  

Plaquette excitations with $b_p=-1$ are also lineons and therefore display dimensional reduction. 
Such excitations are located at the open end of a diagonal in the pictorial representation. 
If one moves the excitation orthogonal to this diagonal by a local perturbation, 
one always introduces a kink in the diagonal and therefore creates a fracton (see Fig.~\ref{Fig::fractons} bottom right).  
As a consequence, the plaquette excitations can only move along a diagonal or anti-diagonal without creating fractons and are therefore also lineons. 

{\it{Topological phase:}}
The ground state of the XYTC displays $\mathbb{Z}_2$ topological order whenever it is adiabatically connected to the cases $\phi =0$ and $\phi=\pi/2$ where the model reduced to the conventional toric code. 
For any $J_{s_1}\neq \{0,1\}$ one therefore has a topological ground-state degeneracy $4^g$ with $g$ the genus of the underlying topology and a finite topological entanglement entropy quantifying the long-range entanglement of the ground states.
Interestingly, as for the fracton phase, the excitations of this topological phase exhibit restricted mobility similar to the toric code in a transverse field \cite{Vidal_2011} due to the sub-extensive number of conserved parities along (anti-) diagonals.

Let us focus on the limit $\phi=0$ with \mbox{$J_{s_1}=J_{s_2}=1/2$} and consider $\hat{A}_s^{y}$ as a perturbation. 
The star operator $\hat{A}_s^{y}$ flips the eigenvalues $a_{s'}^{(x)}$ with $s'$ the four neighboring stars of $s$ while it does not change any plaquette eigenvalue $b_p$.
As a consequence, configurations with an arbitrary number of flux excitations with $b_p=-1$ are fully immobile.
For the charge excitations with $a_{s'}^{(x)}=-1$ the mobility depends strongly on the number and spatial distribution of charges.
This can be seen best when considering the effects of the conserved parities relative to the ground states having all parities $+1$. 
Indeed, the presence of an odd number of charges along one (anti-) diagonal implies one parity $-1$ along this direction.   
A single charge is therefore immobile because it is fixed by the two odd parities corresponding to one diagonal and one anti-diagonal crossing the star.    
Similarly, two charges on an (anti-) diagonal exhibit a one-dimensional dispersion and only four charges on a rectangular configuration are allowed to move in two dimensions.

{\it{Conclusions:}} In this work we have introduced the XYTC. Its quantum phase diagram can be determined exactly by a duality mapping to two self-dual Xu-Moore models due to the existence of an extensive number of conserves subsystem parities. For any spatial anisotropy of the star sublattices, it displays  two phases with $\mathbb{Z}_2$ topological order and in between a phase with type-I fracton order. The excitations display dimensional reduction in both phases. 

There exist several exciting extensions. 
One can extend the XYTC to three dimensions starting from the conventional $\mathbb{Z}_2$ toric code on the cubic lattice \cite{Hamma_2005, Nussinov_2008, Reiss_2019}, where one expects the appearance of fracton and topological phases due to the existence of planar subsystem parities. 
Next it would be interesting to link our analytic findings to the physical properties of the non-solvable U(1) toric code \cite{Wu_2023}, e.g., by also deforming the U(1) toric code to the U(1) checkerboard toric code and investigating the limit of isolated star sublattices.  
Further, the exploration of potentially continuous quantum phase transitions out of the XYTC fracton phase via lineon condensation is promising.  
Altogether, it would certainly be valuable to generalize our approach for the XYTC to other quantum spin liquids including non-Abelian phases.

{\it{Acknowledgments:}}
We thank Viktor Kott, Lea Lenke, and Andreas Schellenberger for fruitful discussions. KPS gratefully acknowledge financial support by the Deutsche Forschungsgemeinschaft (DFG, German Research Foundation) through the TRR 306 QuCoLiMa ("Quantum Cooperativity of Light and Matter") - Project-ID 429529648 (KPS). KPS acknowledges further financial support by the German Science Foundation (DFG) through the Munich Quantum Valley, which is supported by the Bavarian state government with funds from the Hightech Agenda Bayern Plus.

\bibliography{bibliography_2.bib}

\end{document}